\documentclass{iau}
\usepackage{graphicx}
\usepackage{hyperref}

\title[Close encounters of NEOs] 
{Close encounters of Near Earth Objects\\ with large asteroids}

\author[Anatoliy Ivantsov, Siegfried Eggl, Daniel Hestroffer, William Thuillot \& Pini Gurfil]   
{Anatoliy Ivantsov$^{1,2}$,
Siegfried Eggl$^2$,
Daniel Hestroffer$^2$,
William Thuillot$^2$
 \and Pini Gurfil$^1$}

\affiliation{$^1$Faculty of Aerospace Engineering, Technion - Israel Institute of Technology, \\ Technion City,
IL-3200003, Haifa, Israel \\ email: {\tt ivantsov@tx.technion.ac.il} \\[\affilskip]
$^2$Institut de M{\'e}canique C{\'e}leste et de Calcul des Eph{\'e}m{\'e}rides, Observatoire de Paris, \\ 77 avenue Denfert Rochereau,
F-75014, Paris, France}

\pubyear{2014}
\volume{310}  
\pagerange{119--126}
\jname{Complex Planetary Systems}
\editors{Z. Kne{\v z}evi{\'c}, ed.}
\begin{document}

\maketitle

\begin{abstract}

Close encounters of Near Earth Objects (NEOs) with large asteroids are a possible source of systematic errors in trajectory propagations and asteroid mitigation. 
It is, thus, necessary to identify those large asteroids that have to be considered as perturbers in NEO orbit modeling. 
Using the Standard Dynamical Model we searched for encounters between the 1649~numbered Near Earth Asteroids (NEAs) and 2191~large asteroids having sizes greater than 20~km. 
Investigating the $21^{st}$ century A.D. we have found 791 close encounters with 195 different large asteroids that lead to a substantial scattering of NEOs.
\keywords{Near-Earth objects, asteroids, ephemerides}
\end{abstract}

\firstsection 
\section{Introduction}
Close encounters of NEAs with the nearby planets and the Moon are known to be the source of strong
gravitational perturbations on the trajectories of the former. In contrast, the influence of massive
asteroids has not received a lot of attention. Yet, close encounters with massive asteroids can 
be the sources of systematic errors in trajectory propagation, asteroid mitigation and sample return missions,
see e.g.  \cite[Eggl et al. (2013)]{Eggl13}, \cite[Chesley et al. (2014)]{Chesley14} and references therein.
The purpose of the current research is to identify those massive asteroids that should be considered in the trajectory
propagation of known NEAs. 

\firstsection 

\section{Dynamical Model and Massive Asteroids}
The Standard Dynamical Model for orbit propagation of asteroids introduced by
\cite[Giorgini et al. (2008)]{Giorgini08} includes all relativistic gravitational forces caused by the Sun,
the planets, the Moon, as well as Ceres, Pallas and Vesta. However, the present research differs in one important
aspect from the standard model.

Since NEAs can have close encounters not only with the Earth, but also with the Moon, Venus and
Mars, the additional forces due to asphericity of these bodies should be taken into consideration
in some cases. The PHA (99942) Apophis, for instance, will have
encounters closer than 0.1~au with other bodies than the Earth within the next 15~years. In fact, it comes close to Venus
on April~24, 2016  (distance of 0.0782~au) and the Moon on April~14, 2029 (distance of 0.00064~au).
For this reason we argue that it is necessary to include the influence of the $J_2$ gravitational
harmonic for the Earth, the Moon, Venus, and Mars whenever a NEA appeared within the gravitational
sphere of influence of one of the above mentioned bodies. The Sun's $J_2$ gravitational harmonic was
considered as well. 

Cross-referencing asteroid diameters (D) listed in the NEOWISE project \cite[Mainzer et al. (2011)]{Mainzer11} with JPL's Small-Body
Database lead to a preselection of 2191 massive asteroids with D$>$20~km
that were used as potential perturbers in our simulations.
The bulk-density adopted for the
calculation of gravitational parameters for the perturbing asteroids was 3~g/cm$^3$. 

All asteroid trajectories were propagated using an Adams-Moulton integrator of variable step and
order, while the positions and velocities of all the major bodies were interpolated from the
recent JPL DE430 ephemerides. Initial conditions for the asteroids were taken from the JPL
HORIZONS system. The equations of motion were integrated for a century starting from J2000.
\firstsection
\section{Close Encounter Detection}
We searched for those close encounters between the preselected asteroids greater than 20~km in size
and all known 1649~NEAs that would result in a two-body deflection angle greater than 0.1~arcsec. Table \ref{tab1}
contains a small sample of the encounter parameters and observational conditions for specific close encounters
between NEAs and perturbing asteroids. The visual deflection angle represents the deflection as visible from
the Earth, thus allowing to estimate whether the deflection event will be or has been observable.

\begin{table}
  \begin{center}
  \caption{Sample from the list of closest encounters of NEOs with large asteroids.$^1$}
  \label{tab1}
 {\scriptsize
  \begin{tabular}{cccccccccc}\hline 
{\bf Perturbing} & {\bf NEA} & {\bf JD} & {\bf Calendar} & {\bf Minimum} & {\bf Relative} & {\bf Hill's} & {\bf Deflection,} & {\bf Visual} & {\bf Solar} \\ 
{\bf asteroid} & & &{\bf date} &{\bf distance,} &{\bf speed,}&{\bf radius}&{\bf arcsec} & {\bf deflect.,}&{\bf elong.,} \\
&&&&{\bf au}&{\bf au/day}&&&{\bf arcsec}&{\bf deg.} \\ \hline
1 & 8013 & 2454388 & 2007-10-14 & 0.02178 & 0.004119 & 12.6 & 0.225 & 0.185 & 147 \\
52 & 184266 & 2458552 & 2019-03-09 & 0.00292 & 0.003119 & 4.1 &0.127 & 0.057 & 13 \\
1 & 3551 & 2461874 & 2028-04-12 & 0.02554 & 0.005619 & 15.7 & 0.103 & 0.090 & 150 \\
  \end{tabular}
 }
 \end{center}
\vspace{1mm}
 \scriptsize{
 {\it Notes:}\\
  $^1$The complete list encompasses 791 encounters between the numbered NEOs and 195 massive perturbing asteroids and is available \url{http://www.imcce.fr/~ivantsov/nea21cy.txt}}
\end{table}
\firstsection
\section{Conclusions}
An accurate orbit prediction of NEOs requires a more detailed gravitational
interactions than suggested in the Standard Dynamical Model. In the case of close encounters between NEOs and
the nearby planets as well as the Moon planetary asphericities should be considered. Some encounters with larger
asteroids are also non-negligible. Including only the three largest bodies in the main
belt (Ceres, Pallas, Vesta) may lead to inaccurate orbit predictions.  
The gravitational effect of 195 main belt asteroids should be taken into account for the current NEO population. Rectifying the dynamical model is  
especially important for the orbital prediction of Potentially Hazardous Asteroids, such as (99942) Apophis.
In a next step we will investigate the robustness of our results with respect to non-gravitational forces.
\\
\\
\textit{Acknowledgments:} S.E. was supported by the European Union Seventh Framework Program under grant agreement
no. 282703 and the IAU Symposium no. 310 travel grant. 

\firstsection

\end{document}